# Backscattering immunity of dipole-exchange magnetostatic surface spin waves

M. Mohseni,[1] R. Verba,[2] T. Brächer,[1] Q. Wang,[1] D. A. Bozhko,[1] B. Hillebrands,[1] and P. Pirro[1,*]

[1] *Fachbereich Physik and Landesforschungszentrum OPTIMAS, Technische Universität Kaiserslautern, 67663 Kaiserslautern, Germany.*

[2] *Institute of Magnetism, Kyiv 03680, Ukraine.*

The existence of backscattering-immune spin-wave modes is demonstrated in magnetic thin films of nano-scale thickness. Our results reveal that chiral Magneto Static Surface Waves (cMSSWs), which propagate perpendicular to the magnetization direction in an in-plane magnetized thin film, are robust against backscattering from surface defects. cMSSWs are protected against various types of surface inhomogeneities and defects as long as their frequency lies inside the gap of the volume modes. Our explanation is independent of the topology of the modes and predicts that this robustness is a consequence of symmetry breaking of the dynamic magnetic fields of cMSSWs due to the off-diagonal part of the dipolar interaction tensor, which is present both for long- (dipole dominated) and short-wavelength (exchange dominated) spin waves. Micromagnetic simulations confirm the robust character of the cMSSWs. Our results open a new direction in designing highly efficient magnonic logic elements and devices employing cMSSWs in nano-scale thin films.

Protected transport of energy and information has gained a tremendous amount of interest during the last decade [1–4]. Magnons, the quanta of spin waves (SWs), which are the collective excitations of the spin ensemble of a magnetically ordered system, are considered as a promising counterpart to photons and phonons to serve as information carriers in future wave-based data processing devices [5–19]. For the design of magnon-based devices, decreasing propagation losses is considered as one of the biggest challenges [18,20]. In addition to intrinsic magnetic losses manifesting themselves in a viscous damping, external losses mediated by surface defects such as fabrication-induced disorders, roughness and magnetic inhomogeneities, generally contribute to the total losses. Therefore, novel ways to avoid these scattering losses are highly desired.

In fact, this has motivated a large number of theoretical works, e.g., on the potential topological protection of SWs [21–26]. Nevertheless, the observation of protected magnon transport remains still a great challenge since most of the proposed systems obtain their topological protection from Dzyaloshinskii-Moriya interaction (DMI) [27–30] or strongly inhomogeneous magnetic ground states [21]. Such properties are hard to realize experimentally or can come along with a detrimental influence on the propagation properties [28,30]. Considering these facts, the prediction of protected magnon transport which does not rely on topological arguments would constitute a major breakthrough [20].

Here, we introduce a mechanism of wave backscattering immunity which is distinct to that topological insulators. We show that backscattering immune SW modes exist in simple thin film systems which have homogeneous magnetic parameters and which do not exhibit DMI. In terms of applications, the studied system benefits largely from its simplicity in comparison to artificially created metamaterials and crystals for robust photonic [31,32], phononic [33] and magnonic transport. We show that the protection of these modes is caused by the chirality of the mode profiles of the counter-propagating SWs. In particular, we present a theory suggesting that chiral Magneto Static Surface Waves (cMSSWs) are robust against backscattering from surface defects. We further show that this protection is effective for both dipolar waves as well as for short-wavelength waves dominated by the exchange interaction. Using micromagnetic simulations, we confirm the predictions of our theory for an Yttrium Iron Garnet (YIG) model system which is the most suitable host for SW propagation due to its low intrinsic losses [6,8,9,12,13,18,34–38]. The obtained results can be generalized to other materials like metallic alloys. The simulations also show that the protection against scattering is particularly strong if the cMSSW frequency is located in the frequency gap of the volume modes (VM, a.k.a. Perpendicular Standing Spin Waves, PSSW) which opens due to the quantized exchange energy in thin films. Therefore, the backscattering protection is more pronounced in thinner films, in which the gap of VMs is larger.

---

[*] ppirro@rhrk.uni-kl.de

To illustrate the mechanism which leads to the chiral protection of cMSSW, Fig. 1a exemplarily shows its mode profiles and dispersion relations in a YIG film of thickness $d = 80$ nm for typical values of the exchange constant $A_{\text{exch}} = 3.5$ pJ/m and saturation magnetisation $M_s = 140$ kA/m. The case of cMSSWs is realized when the film is magnetized in-plane, perpendicularly to the wave propagation direction, $\vec{M_0} \perp \vec{k}$ (Fig. 1a1). This "Damon-Eshbach" geometry is well studied in the case of pure dipolar (magnetostatic) SWs in relatively thick films. SWs in this geometry are chiral due to the off-diagonal part of the dynamic dipole-dipole interaction, which breaks time-reversal symmetry ("$T$-symmetry"). This, inter alia, leads to a non-reciprocal localization of cMSSWs at the surfaces of the film which depends on the direction of the wave vector $k_y$ [38–40]. Considering the nano-scale thickness of the system studied here, both dipolar and exchange interactions are important for the wave propagation.

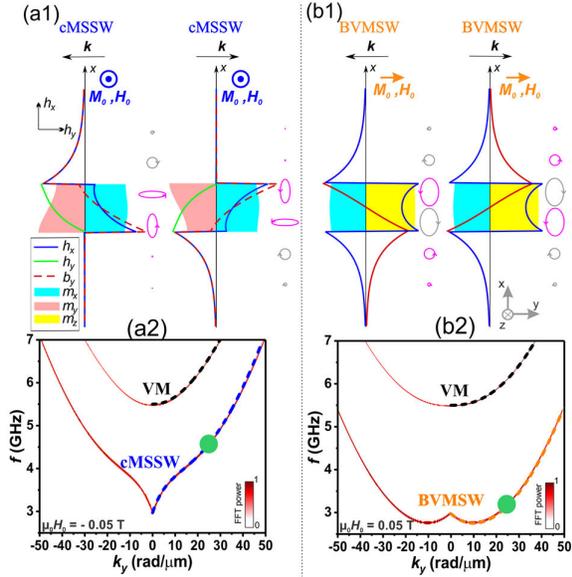

Figure 1. (a1, b1). Profiles of cMSSW and BVMSW, propagating in opposite directions (color coded), and corresponding distributions of the dynamic magnetic field **h** and the induction **b**. Ellipses show the vector structure of the dynamic magnetic field **h**. The color code indicates the sign of rotation of the dynamic magnetic field. Film thickness 80 nm and $k_y = 25.9$ rad/µm. (a2, b2) Corresponding magnon band structures are calculated via numerical simulations (color plot) and analytical calculations according to [41] (dashed lines). The fundamental modes are distinguished via blue (cMSSW) and orange (BVMSW) lines, and the higher order volume modes (VM) via black lines. Green dots indicate the modes exemplarily investigated in Fig. 2.

Under these conditions, the dynamic magnetization components of cMSSWs are weakly localized at one of the surfaces (see Fig. 1a). As it is also visible from the dispersion relations shown in Fig. 1a2, higher-order volume modes (VMs) are present in addition to cMSSW. The exchange energy contribution, which is quantized over the thickness with a quantum proportional to $1/d^2$, leads to a frequency shift of VMs above the cMSSW frequency. This implies a frequency gap where only the cMSSW is present.

For comparison, we also study the case of a non-chiral wave (Fig. 1b): if the wave vector is parallel to the static magnetization ($\vec{k} \parallel \vec{M_0}$), so-called Backward Volume Magneto Static Waves (BVMSW) occur. For this orientation, the dipolar interaction does not lead to broken T-symmetry.

As we already pointed out, the spatial localization of cMSSW at a film surface is weak in the studied systems. To show how the chirality nevertheless leads to backscattering protection, the corresponding equations of motion are analysed. For the appearance of any chiral and nonreciprocal effects, at least the $T$-symmetry should be broken. In principle, magnetization dynamics always exhibits broken $T$-symmetry, since magnetization precesses always counter-clockwise around the static field direction. However, this is not a sufficient criterion. Indeed, the propagation of small-amplitude SWs is described by the following dynamical equations [41-42]:

$$\frac{dm_x}{dt} = -\left(\omega_H - \omega_M \lambda^2 \nabla^2\right) m_y - \omega_M b_y$$
$$\frac{dm_y}{dt} = \left(\omega_H - \omega_M \lambda^2 \nabla^2\right) m_x + \omega_M b_x \quad (1)$$

where $\omega_H = \gamma B$, $B$ is the static internal field, $\omega_M = \gamma \mu_0 M_s$, $\lambda$ is the exchange length, $b(r) = \int \hat{G}(r,r') \cdot m(r') dr'$ is the dynamic dipolar field with $\hat{G}$ being the magnetostatic Green function, and the static magnetization is assumed to be oriented along the $z$-direction. If the off-diagonal part of dynamic interaction is zero $G_{xy} = G_{yx} = 0$, the equation system (1) can be transformed to one equation for $m_x$ (or $m_y$), which contains *only* second-order time derivatives of $m_x$, i.e., of the form $d^2 m_x(r,t)/dt^2 = F[m_x(r,t),r]$ with $F$ being the integro-differential operator, which does not depend explicitly on the time $t$. In this scenario, SW propagation exhibits $T$-symmetry and does not possess chirality by itself. This case is realized for BVMSW, as for these waves, the dynamical components of magnetization ($z$ and $x$ in notation of Fig. 1) are not dipolarly coupled.

In contrast, in the cMSSW geometry, the dynamic magnetization components are coupled and $G_{xy} = G_{yx} \neq 0$. Consequently, Eq. (1) cannot be simplified to one equation that contains only even time derivatives of magnetization. Hence, $T$-symmetry of cMSSW propagation is broken and these waves are chiral.

To understand how this chirality leads to protection against backscattering, one first has to note that the non-zero off-diagonal components of the dynamic dipolar interaction result in a completely different symmetry for the dipolar fields created by cMSSW and BVMSW (see Fig. 1). The dipolar fields have to satisfy the electromagnetic boundary conditions at any internal or external boundary like the boundary of a defect. Namely, this implies continuity of the tangential component of the magnetic field $h$ and of the normal component of the magnetic induction $b = \mu_0 (h + m)$. In addition, the exchange interaction requires continuity of magnetization and its spatial derivative. The resulting different form of the created fields has direct implications on the reflection and transmission of the waves.

In the case of BVMSW, the dipolar fields for opposite propagation directions are related by mirroring with respect to the X0Z plane, and the spatial distribution of the magnitude of the fields is the same (see Fig. 1b). In this sense, forward and reflected waves match each other, and direct reflection of BVMSW into an oppositely propagating wave can easily occur satisfying the boundary conditions. In contrast, dipolar fields of oppositely propagating cMSSWs are related by subsequent mirroring to the Y0Z and the X0Z plane, respectively. This is also sketched with the ellipses in Fig. 1, showing the precession of the dynamic magnetic field $h$. Consequently, the field strength at a given vertical position differs significantly. This can, for instance, be seen in the regions above and below the film. This feature is due to the constructive and destructive interference of the dipolar fields produced by the $x$- and $y$- components of magnetization. In particular, this feature is still present even in the case when the dynamic magnetization distribution in the film is uniform. Thus, dipolar fields of forward and backscattered cMSSWs are "incompatible" and cMSSW cannot simply scatter to oppositely propagating cMSSW because of the impossibility to satisfy the boundary conditions.

In order to shed more light on this "mismatch" of the cMSSW profiles we consider the following simple model. We assume, that only propagating cMSSWs are involved in the dynamics. Furthermore, we represent the magnetic fields at the boundary as the sum of the fields created by incident cMSSW, transmitted cMSSW with amplitude $T$, and reflected cMSSW with amplitude $R$. The regions before and after the boundary situated at $y = y_g$ are different only by their thickness. This corresponds, for instance, to an unperturbed film in contact with a defect, as is shown in Fig. 2. As was pointed out above, with propagating cMSSWs only, it is impossible to exactly satisfy the boundary conditions. In practice, other localized evanescent waves are involved in the scattering process. Nevertheless, we can find values of the transmission and reflection coefficients which minimize the mismatch of the fields at the boundary. The obtained values cannot be interpreted as real transmission and reflection coefficients, but they show the qualitative behavior of these coefficients in different conditions.

Using the analytically calculated profiles of cMSSWs, we construct the functional of integral mismatch of the field $h_x$ at the boundary at the position $y = y_g$:

$$\Phi_h(R,T) = \int_{-\infty}^{+\infty} \left| h_{x,k_1} + R h_{x,-k_1} + T h_{x,k_2} \right|^2 dx.$$

Here, $k_1$ and $k_2$ are the wavenumbers of incident and transmitted spin waves, which are related by the dispersion relation $\omega_1(k_1) = \omega_2(k_2)$. The same functional was constructed for the mismatch of the component of the magnetic induction $b_y$. As one cannot directly compare magnetic field and induction, we minimize both functionals separately assuming complex-valued coefficients $R$ and $T$, i.e., allowing for an arbitrary phase shift between waves.

From this, we infer that the minima of mismatches of $b_y$ and $h_x$ take place for approximately the same transmission rate $T$. However, at the same time, the reflection rate $R$ has opposite signs with similar magnitude for the two components. In other words, the conditions of continuity of the normal component of the magnetic induction and the tangential component of the field imply that both act on the reflected wave oppositely, i.e., with a phase shift of $\pi$. This phase relation is a direct consequence of the specific symmetry of the field profiles of the cMSSW and it suppresses the formation of the reflected wave. This leads to almost 100% transmission from the defect. The calculations also show that this feature – opposite signs of $R$ – remains valid for defect depths ranging from a vanishing depth up to depths of about 50 % of the film thickness.

In the following, we will verify the analytical predictions using micromagnetic simulations. The simulations have been performed via the MuMax 3.0 open source software [43]. The simulated film parameters are equivalent to the ones of Fig. 1 and a Gilbert damping parameter of $\alpha = 0.0002$ has been assumed. The external magnetic field of 0.05 T is applied along the $-z$ direction or $+y$ direction (see supplementary for more information).

To check for backscattering immunity, in Fig. 2 a topographical defect (2 μm long), with a height equal to $h = 20\%$ of the film thickness $d$ is placed at one of the surfaces of the film. Snapshots from micromagnetic simulations of propagating SW pulses with a length of 10 ns excited with a SW source at the centre ($y = 0$ µm) are shown in Fig. 2 for three different times ($t_1$ to $t_3$) before and after reaching the defect.

In all presented cases, SWs with a wave vector of $k_y = 25.9$ rad/µm, as indicated by the green dots in Fig. 1, have been studied. For the cMSSW (Fig. 2a) the system is excited with a carrier frequency of $f = 4.64$ GHz. In very good agreement with our prediction, the cMSSW pulse (blue arrow to the left) passes the defect without any significant reflection and reaches an amplitude transmission close to 96 % (see supplementary animation Fig. 2a). However, in the case of the BVMSW, shown in Fig. 2b, the SW pulse undergoes a strong back-reflection when impinging on the defect and only 62 % of the wave is transmitted (see supplementary animation Fig. 2b).

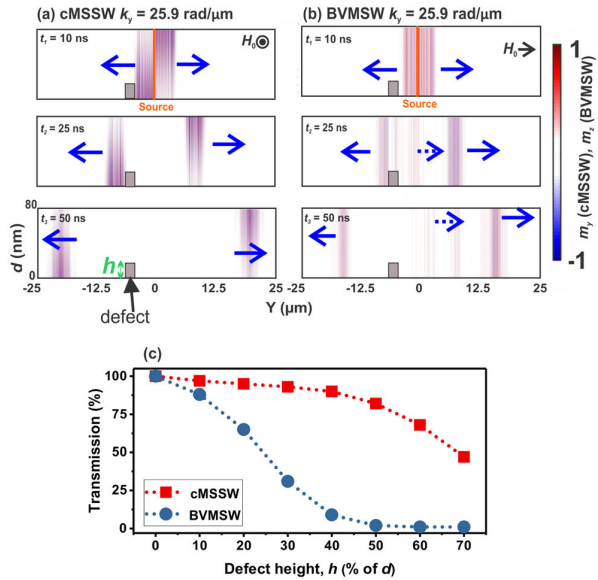

Figure 2. Snapshots of the SW propagation in the presence of a defect (grey rectangular). (a): cMSSW ($f = 4.64$ GHz and $k_y = 25.9$ rad/µm), b: BVMSW ($f = 3.2$ GHz and $k_y = 25.9$ rad/µm). Direction of the applied field $H_0$ are indicated in the right corner. Transmitted and reflected waves are marked with solid and dashed arrows, respectively, c: SW transmission for different SW modes as a function of the topographic defect height. Lines are guides to the eye.

In Fig. 2c, we present a systematic characterization of the scattering from topographical defects for the same parameters as used in Fig 2a-b. We vary the defect height ($h$) as displayed in Fig 2a, and evaluate the transmission of the propagating SWs. It can be seen that for the cMSSW case, a defect can be as high as 40 % of the thickness, and still a transmission in the range of 90 % can be achieved. This is in good agreement with the estimation from the analytical part presented above. A further increase of the defect height creates a coupling channel which allows the wave to scatter to the other surface and propagate backwards. In contrast, for the BVMSW case (Fig 2b), already a defect as high as 40 % of the thickness is enough to result in an almost complete reflection. Here, the scattering takes place between $k_y$ and $-k_y$ of the same mode. In addition, we verified that this strong scattering can be observed over the whole wave vector range shown in Fig. 1b2

In Fig. 3a1 the transmission of cMSSW for the same system as discussed in Fig. 2a is shown as a function of the excited wave vector. From small wave vectors starting from $k_y = 2$ rad/µm (with dominant dipolar interaction) to midrange $k_y = 35$ rad/µm (dipole-exchange waves), the cMSSW transmission is around 95%. In this frequency range (which is marked by the blue coloured area of Fig 3a1), the cMSSW lies inside the gap of the VM. Therefore, no resonant scattering to the VM is possible due to energy conservation.

A drop of the cMSSW transmission by roughly 20 % is visible in Fig. 3a1. This drop appears for wave vectors $k_y > 36$ rad/µm, when the cMSSW becomes frequency-degenerated with the first VM. This degeneracy enables a resonant scattering from the cMSSW to the VM. This supplies a channel for back-reflection of SW energy. The appearance of backscattered VMs is clearly visible in the simulations (see supplementary animation Fig. 3).

It should be noted that the strong protection that is found even for low wave vectors provides evidence that the localization of the dynamic magnetization at a surface plays no direct role for the protection. Indeed, the localization on one surface is proportional to the in-plane wave vector $k_y$ [40], and, e.g., for $k_y = 2$ rad/µm, the amplitude decays by only 4.2 % from one side to the other and the mode profile is nearly homogeneous.

As a next step, we will show that the strong protection of cMSSW in the thin film is a general phenomenon which is present even for high wave vector SWs with a dominant exchange energy contribution. Theoretically, this can be predicted by evaluating the off-diagonal component of the Green function, which is responsible for the specific symmetry of dipolar fields of the cMSSW [41]:

$$G_{xy}(k_y, x, x') = \frac{i}{2}\text{sign}[x-x']k_y e^{-|k_y(x-x')|} \quad (2)$$

The strength of the asymmetry can be naturally measured by the difference of the off-diagonal contribution at the opposite surfaces of the film. These can be estimated as $(1 - \exp[-|k_y d|])$. This term increases with $k_y$, giving insight why cMSSW protection is present even in the high-$k$ exchange-dominated range.

In order to prove micromagnetically that the protection still exist for exchange-dominated SWs, the scattering study of Fig 3a is repeated for a film with $d$ = 30 nm which is shown in Fig. 3b1. Due to the stronger quantization, this system shows a much larger gap of VMs and the protection of shorter wavelength can be tested without the occurrence of the scattering channel to the VMs. Similar to the 80 nm film, the protection exists as long as no resonant scattering to VM is possible. In this case, the maximal wave number at which cMSSW are not degenerated with VMs is $k_y \sim 100$ rad/µm. For these waves, the exchange contribution to the SW energy ($\sim \omega_M \lambda^2 k^2$) is almost 4 times larger than the dipolar one ($\sim \omega_M (1 - e^{-kd})$).

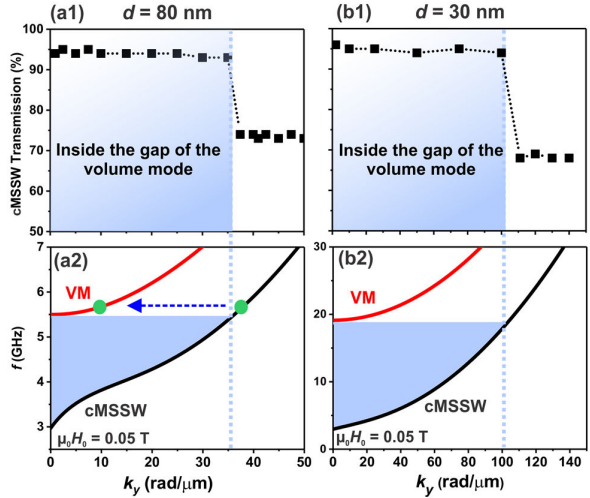

Figure 3. cMSSW transmission as a function of wave vector in the presence of a 2 µm × (20% of $d$) topographical defect for a film with a1: $d$ = 80 nm, and b1: $d$ = 30 nm. The blue area shows the range of wave vectors inside the gap of volume modes where the protection is strong, a2 and b2: indicate the relevant magnon bandstructure of the systems and the corresponding frequency gap which is distinguished via the blue area. Please note the different scales.

In conclusion, we showed that chiral Magneto Static Surface Waves (cMSSWs), which propagate perpendicular to the static magnetization in an in-plane magnetized thin film, are robust against backscattering from surface defects. The protection of the cMSSW can be understood without a consideration of the topology of the system. It is strong if the-frequency lies inside the gap of the volume modes, where no resonant scattering to or hybridization with other modes is possible. It should be emphasized, that this protection takes place in ferromagnetic films with nano-scale thicknesses both for the dipole-dominated and the dipole-exchange range. It is also observed that the localization of the magnetization profile of cMSSW is not the protecting mechanism from back-scattering. At the same time, it is also observed for exchange dominated waves as long as they lie in the gap of the volume modes. Absolute protection of the cMSSWs is expected if the inversion symmetry of the system is broken, e.g. using a bilayer to design the magnon band structure and change the mode profiles of the counter-propagating SWs. This knowledge gives a new view on the role of the dipolar interaction on exchange dominated SWs. The thin film system proposed here is the simplest medium for protected magnon transport and probably, in general, protected energy transport via waves.

Financial support by the Deutsche Forschungsgemeinschaft (SFB/TRR 173 "Spin+X", Project B01), by the Nachwuchsring of the TU Kaiserslautern and by the European Research Council Starting Grant 678309 MagnonCircuits is gratefully acknowledged. R.V. acknowledges support from Ministry of Education and Sciences of Ukraine (project 0118U004007).

The authors would like to thank K. Yamamoto for fruitful discussions about the topological properties of the MSSWs.